\documentclass[12pt]{article}
\usepackage{graphicx}
\usepackage{amssymb}
\usepackage{amscd}
\usepackage{amsmath}
\usepackage{appendix}
\usepackage{cite}

\begin{document}
\begin{titlepage}

{\hbox to\hsize{\hfill February 2017 }}

\bigskip \vspace{3\baselineskip}

\begin{center}
{\bf \large 

Electroweak monopoles and the electroweak phase transition}

\bigskip

\bigskip

{\bf Suntharan Arunasalam and Archil Kobakhidze  \\ }

\smallskip

{ \small \it
ARC Centre of Excellence for Particle Physics at the Terascale, \\
School of Physics, The University of Sydney, NSW 2006, Australia \\
E-mails: suntharan.arunasalam, archil.kobakhidze @sydney.edu.au
\\}

\bigskip
 
\bigskip

\bigskip

{\large \bf Abstract}

\end{center}
\noindent 
We consider an isolated electroweak monopole solution within the Standard Model with a non-linear Born-Infeld extension of the hypercharge gauge field. Monopole (and dyon) solutions in such an extension are regular and their masses are predicted to be proportional to the Born-Infeld mass parameter. We argue that cosmological production of electroweak monopoles may delay the electroweak phase transition and make it more strongly first order for monopole masses $M\gtrsim 9.3\cdot 10^3$ TeV, while the nucleosynthesis constraints on the abundance of relic monopoles impose the bound $M\lesssim 2.3\cdot 10^4$ TeV. The monopoles with a mass in this shallow range may be responsible for the dynamical generation of the matter-antimatter asymmetry during the electroweak phase transition.      
 
\end{titlepage}

\baselineskip=16pt

\section{Introduction}

For a long time, there was a prevailing  view that topologically stable monopole solutions do not exist in the Standard Model because the vacuum manifold $SU(2)\times U(1)_Y/U(1)_{EM}$ allows no non-trivial second homotopy group. This has been questioned in 
\cite{Cho:1996qd}, where topological stable monopole (and dyon) solutions, representing a non-trivial hybrid between $U(1)_{EM}$ Dirac monopole \cite{Dirac:1948um} and non-Abelian 't Hooft-Polyakov monopole \cite{'tHooft:1974qc, Polyakov:1974ek}, have been found in \cite{Cho:1996qd}. While the SU(2) non-Abelian configuration is regular, the $U(1)_{EM}$ configuration exhibits point singularity at the origin. As a result, the monopole mass is divergent. There is no obvious problem with the energy of a classical configuration being divergent as it may be regularised in a more complete quantum theory. In fact, some regularized monopole solutions have been also proposed \cite{Bae:2002bm, Cho:2013vba, Cho:2012bq, Ellis:2016glu}, which indicate that electroweak monopoles as light as $\sim 5-10$ TeV may actually exist. 

In this paper, we consider the Standard Model where the standard kinetic term for $U_Y(1)$ hypercharge gauge boson 
is a part of a non-linear Born-Infeld Lagrangian. In this theory we account for an extra mass parameter, $\sqrt{\beta}$, which controls the non-linearity of the  hypercharge field. Similar to the regularisation of the divergent energy of a point-like charge in the original Born-Infeld electrodynamics \cite{Born:1934gh}, we find that the electroweak monopole gets also regularised and it mass predicted to be $\propto \sqrt{\beta}$.     

The electroweak monopoles must be copiously produced during the electroweak phase transition via the Kibble mechanism \cite{Kibble:1976sj, Preskill:1979zi}\footnote{Strictly speaking the Kibble mechanism is applicable to global monopole production. The refined mechanism in the case of gauge theories is discussed in \cite{Rajantie:2002dw}.}. Furthermore, as we will argue in this paper, cosmological production of electroweak monopoles may delay the electroweak phase transition and make it stronger first order. The physics behind this can be heuristically explained as follows.  Magnetic monopoles (and antimonopoles) with symmetric vacuum configuration within the monopole core are trapped in the region surrounded by the domains with symmetry-breaking vacua with different orientation of the Higgs field in the $SU(2)\times U(1)_Y/U(1)_{EM}$ vacuum manifold. This costs in energy, leading to a higher free energy in the broken phase relative to the case without monopole production. In particular, we find that $\phi_c/T_c\gtrsim 1$ can be achieved without violating nucleosynthesis constraints on relic monopole abundance.  This may have important implications for electroweak baryogenesis. Namely, sphaleron mediated $B+L$- violating processes become ineffective below the critical temperature, $T_C$, preventing the washout of previously generated baryon asymmetry. 

The rest of the paper is organised as follows. In Section 2, we discuss the electroweak monopole solution within the Born-Infeld hypercharge extension of the Standard Model. Section 3 is devoted to the discussion of the monopole production and its impact on the electroweak phase transition. In section 4, we conclude.

\section{Electroweak monopoles in the Born-Infeld hypercharge model}
Let us consider Standard Model extended by the non-linear Born-Infeld type hypercharge gauge field. The relevant bosonic Lagrangian reads:
\begin{align}
\mathcal{L}&= |D_\mu H|^2-\frac{\lambda}{2}\left(H^\dagger H-\frac{\mu^2}{\lambda}\right)^2-\frac{1}{4}F^i_{\mu\nu}F^{i\mu\nu}+\beta^2\left[1-\sqrt{-\det\left(\eta_{\mu\nu}+\frac{1}{\beta}B_{\mu\nu}\right)}\right] \nonumber \\
\begin{split}
 &=-|D_\mu H|^2-\frac{\lambda}{2}\left(H^\dagger H-\frac{\mu^2}{\lambda}\right)^2-\frac{1}{4}F_{\mu\nu}F^{\mu\nu}\\
 &\ \ \ \ \ +\beta^2\left[1-\sqrt{1+\frac{1}{2\beta^2}B_{\mu\nu}B^{\mu\nu}-
 	\frac{1}{16\beta^4}(B_{\mu\nu}\tilde{B}^{\mu\nu})^2}\right]
\end{split}
\end{align} 
where $D_{\mu}=\partial_\mu-i\frac{g}{2}\tau^a A^a_{\mu}-i \frac{g'}{2}B_\mu$ is the $SU(2)_L\times U(1)_Y$ gauge covariant derivative with  $A^a_{\mu}$ and $B_{\mu}$ being $SU(2)_L$ and $U(1)_Y$ gauge vector fields, respectively and $H$ is the electroweak doublet Higgs field. $F_{\mu\nu}^a$ ($a=1,2,3$) denote the $SU(2)_L$ gauge field strength tensors, and $\tilde{B}^{\mu\nu}=\frac{1}{2}\epsilon^{\mu\nu\alpha\beta}B_{\alpha\beta}$ is the Hodge  dual to the $U(1)_Y$ field strength tensor $B_{\mu\nu}$. Parameter $\beta$ is a new Born-Infeld parameter of dimension mass$^2$. It controls non-linearity of the hypercharge field and, as we will see shortly, provides ultraviolet regularization of the electroweak monopole mass. In the limit $\beta\to \infty$, we recover the Standard Model theory.

The above Lagrangian leads to the following set of field equations of motion:
\begin{align}
D_{\mu}(D_{\mu}H)&=\lambda\left(H^\dagger H-\frac{\mu^2}{\lambda}\right)H~,
\label{eom1}\\
\left(\partial_\mu-i\frac{g}{2}\tau^a A^a_{\mu}\right)F_{\mu\nu}^i&=i\frac{g}{2}\left[H^\dagger\tau^i(D_{\nu}H)-(D_{\nu}H)^\dagger \tau^i H\right]~,
\label{eom2}\\
\partial_{\mu}\left[\frac{B^{\mu\nu}-\frac{1}{4\beta^2}\left(B_{\alpha\beta}\tilde{B}^{\alpha\beta}\right)\tilde{B}^{\mu\nu}}{\sqrt{1+\frac{1}{2\beta^2}B_{\alpha\beta}B^{\alpha\beta}-	\frac{1}{16\beta^4}(B_{\alpha\beta}\tilde{B}^{\alpha\beta})^2}}\right]&=i\frac{g'}{2}\left[H^\dagger(D_{\nu}H)-(D_{\nu}H)^\dagger H\right]~.
\label{eom3}
\end{align} 
Now, as done in \cite{Cho:1996qd}, consider the following ansatz:
\begin{align}
H&=\frac{1}{\sqrt{2}}\rho\xi, ~\rho=\rho(r),~\xi=i\begin{pmatrix}
\sin(\theta/2)e^{-i\varphi}\\-\cos(\theta/2)
\end{pmatrix}\nonumber\\
A_\mu&=\frac{1}{g}A(r)\partial_\mu t\hat{r}+\frac{1}{g}(f(r)-1)\hat{r}\times \partial_\mu \hat{r}\nonumber\\
B_\mu&=-\frac{1}{g'}B(r)\partial_\mu t-\frac{1}{g'}(1-\cos\theta)\partial_\mu\varphi
\end{align}
In particular, the functions, $A(r)$ and $B(r)$ represent dyon solutions of this model. For $A(r)=B(r)=0$, one obtains pure magnetic monopole, which is also the lightest object and thus we concentrate on this solution. For $A(r)=B(r)=0$, Eq. (\ref{eom3}) is trivially satisfied and Eqs (\ref{eom1}) and (\ref{eom2}) yield:
\begin{align}
\ddot{\rho}+\frac{2}{r}\dot{\rho}-\frac{f^2}{2r^2}\rho&=\lambda \left(\frac{\rho^2}{2}-\frac{\mu^2}{\lambda}\right)\rho\\
\ddot{f}-\frac{f^2-1}{r^2}f&=\frac{g^2}{4}\rho^2f
\end{align}
The following boundary conditions can be chosen for these equations:
\begin{align}
f(0)=1, ~ \rho(0)=0,~ f(\infty)=0, ~\rho(\infty)=\rho_0=\sqrt{\frac{2\mu^2}{\lambda}}
\end{align}
Under these boundary conditions, it can be seen that near the origin,
\begin{align*}
f\approx 1+\alpha_1 r^2,~\rho\approx \beta_1 r^\delta
\end{align*}
with $\delta=(-1+\sqrt{3})/2$ and asymptotically, 
\begin{align*}
f\approx f_1 \exp\left(\frac{-g\rho_0}{2}r\right),~
\rho\approx \rho_0+\rho_1\frac{\exp (-\sqrt{2}\mu r)}{r}.
\end{align*}
The energy of this monopole is given by: 
\begin{align}
E&=E_0+E_1\\
E_0&=\int_0^\infty dr \beta^2\left[\sqrt{(4\pi r^2)^2+\frac{h_Y^2}{\beta^2}}-4\pi r^2\right]\\
E_1&=4\pi\int_{0}^\infty dr \left(\frac{1}{g^2}\frac{(f^2-1)^2}{2r^2}+\frac{1}{2}(r\dot{\rho})^2+\frac{1}{g^2}\dot{f}^2+\frac{\lambda r^2}{8}\left(\rho^2-\rho_0^2\right)^2+\frac{1}{4}f^2\rho^2\right)
\end{align}
where $h_Y=\frac{4\pi}{g'}$ is the hypermagnetic charge of the monopole, $g'$ being the hypercharge gauge coupling. 

Here, $E_0$ is the term corresponding to Born-Infeld hypercharge term and $E_1$ is due to the remainder of the Lagrangian.  In the usual standard model, $E_1$ is finite due to the above boundary conditions and asymptotics and $E_0$ is infinite.   However, due to the Born-Infeld modification, $E_0$ is also made finite.  $E_1$ has been calculated by \cite{Cho:2013vba} to be roughly 4 TeV.  As discussed below, the  mass of the monopoles that  provide a significant impact on the electroweak phase transition must be at least of order $10^4$ TeV.  Hence, $E_0$ must dominate this mass and hence, it is assumed that $E\approx E_0$. This term can be calculated exactly using elliptic integrals as \cite{Kim:1999bd}:

\begin{align}
E&\approx \frac{\pi^{3/2}}{3\Gamma\left(\frac{3}{4}\right)^2} \sqrt{\frac{\beta h_Y^3}{4\pi}} =
\frac{4\pi^{5/2}}{3\Gamma\left(\frac{3}{4}\right)^2}\sqrt{\frac{\beta}{g'^3}}\approx 72.8 \sqrt{\beta}~,
\label{monmass}
 \end{align}
 where we have used $g'=0.357$. Thus the monopole mass is proportional to the Born-Infeld mass parameter, $\sqrt{\beta}$. One can verify that the magnetic charge of this monopole solution is $h=\frac{4\pi}{e}$.  

In the perturbative expansion of the Born-Infeld Lagrangian, which is  valid for low hypercharge field strength, $|B_{\mu\nu}|<\beta$, the lowest order Born-Infeld correction appears as operators of mass dimension 8. They involve only hypercharge field and are suppressed by a factor$ \propto \beta^{-2}$. The best bound on the Born-Infeld mass parameter can be inferred from the PVLAS measurements of nonlinearity in light propagation \cite{DellaValle:2014xoa} (see also \cite{NLQED}):   
\begin{align}
\sqrt{\beta}\gtrsim 5.0\cdot 10^{-4}~{\rm GeV}.
\label{monmass}
 \end{align}
This is clearly a very weak constraint compared to constraints from direct searches of massive monopoles \cite{Acharya:2016ukt}, which in our case implies $\sqrt{\beta}\gtrsim 15.1$ GeV. In contrast, the monopole mass is regularized in \cite{Cho:2013vba, Cho:2012bq} by non-renormalisable operators with mass dimension $n>8+2\sqrt{3}$ operator which also involve the Higgs field. These operators are significantly constrained by LHC data on the Higgs-to-2$\gamma$ decay \cite{Ellis:2016glu}.

\section{Monopole production  and electroweak phase transition}

Consider the one loop high temperature effective potential:
\begin{align}
V(\phi,T)&=D(T^2-T_0^2)\phi^2-ET\phi^3-\frac{1}{4}\lambda_T \phi^4
\end{align}
where, the parameters are defined as
\begin{align}
D&=\frac{1}{8v^2}(2m_W^2+m_Z^2+2m_t^2)\nonumber\\ 
E&=\frac{1}{4\pi v^3}(2m_W^3+m_Z^3)\nonumber\\
T_0&=\frac{1}{2D}(\mu^2-4Bv^2)\nonumber\\
\lambda_T&=\lambda-\frac{3}{16\pi^2v^4}\lambda(m,T)\nonumber\\
\lambda(m,T)&=2m_W^4\ln\frac{m_W^2}{a_b T^2}+m_Z^4\ln\frac{m_Z^2}{a_b T^2}-4m_t^4\ln\frac{m_t^2}{a_f T^2}
	 \end{align}
	 with $\ln a_f=1.14$, $\ln a_b=3.91$, $m_H=125$GeV, $m_W=80.2$GeV, $m_Z=91.2$GeV, $m_t=173$GeV and $v=246$GeV.\\
	 
	 Let $\phi_c(T)$ be the value of the Higgs field at the second minimum. If monopoles are not produced, the Gibbs free energy of the unbroken and broken phases are simply the value of the potential at $\phi=0$ and $\phi=\phi_c(T)$ and the critical temperature, $T_c$ is defined as the temperature at which these are equal. In order to avoid the sphaleron washout constraint, one requires that (e.g. see \cite{Trodden:1998ym} and the references therein). 
	 \begin{align}
	 \frac{\phi_c(T_c)}{T_c}\gtrsim 1. \label{sphaleron}
	 \end{align}
	 With the standard model parameters, this ratio is roughly 0.17, implying that sphaleron mediated processes will washout any pre-existing matter-antimatter asymmetry in the universe. This conclusion is altered significantly once the electroweak phase transition is supplemented by the production of the electroweak monopoles. 

Although the electroweak monopoles are typically heavier that the critical temperature, $M>>T_c$, they are produced during the phase transition when the Higgs field becomes frozen in the broken phase. As a result, there is a finite distance over which, the field is correlated with itself. At distances larger than the correlation length, the Higgs field may point in different directions in the manifold of degenerate vacua . Following Kibble \cite{Kibble:1976sj}, one can argue that a certain density of monopoles is to be expected on this account alone. Monopoles (like vortices in a superconductor) can be thought of as a measure of the disorder remaining in the system, where symmetric (normal) regions trapped by flux quantization in the broken (superconducting) ground state.	Hence, production of monopoles will drive the surrounding plasma out of equilibrium. The equilibrium will be eventually restored once the monopole/antimonopole density will drop due to the monopole - antimonopole annihilation. 

The production of monopoles during the electroweak phase transition offers a qualitatively new picture of baryogenesis at the electroweak scale. In addition to sphalerons, there is additional source of anomalous $B+L$ violation through matter-monopole scattering. These scatterings are known to be unsuppressed \cite{Rubakov:1982fp}, and thus are potentially rapid to contribute to the generation of baryon asymmetry at around $T_c$. Once the equilibrium is achieved however below $T_c$,  sphaleron and monopole mediated processes must become irrelevant in the broken phase. The sphalerons are ineffective if condition (\ref{sphaleron}) is satisfied, while matter-monopole scatterings must decouple once the monopole density becomes low enough due to the monopole - antimonopole annihilation. The later process is also important to not upset the standard Big Bang nucleosynthesis.  In what follows we will concern with washout issues, postponing the full discussion of baryogenesis for future work.   
	
\subsection{Circumventing the sphaleron induced washout of baryon asymmetry}	    
Since monopoles are assumed to be heavier than the critical temperature $T_c$, they can be treated as nonrelativistic point-like particles. 
Monopole - antimonopole interactions, as well as interactions of monopoles with charged particles of plasma, are due to the long-range electromagnetic forces. The monopole - antimonopole annihilation cross section can be approximate as $\sigma_{M\bar M}=d_c^2$, where $d_c=h^2/4\pi T$ is the Coulomb capture distance and $h=4\pi/e$ is the monopole charge. Similarly, a relativistic charged particle with a charge $q$ will scatter off the monopole (antimonopole) with the cross section $\sigma_{qM}=(qh/4\pi)^2 T^{-2}$. The initial density of isolated monopoles (and antimonopoles) can be estimated as $n_0=d^{-3}_c$  \cite{Preskill:1979zi}. Below the capture distance monopole - antimonopole pair would form a unstable bound state and decay subsequently through annihilation.
 
\begin{figure}[t]
\begin{center}	 	
	 	\includegraphics{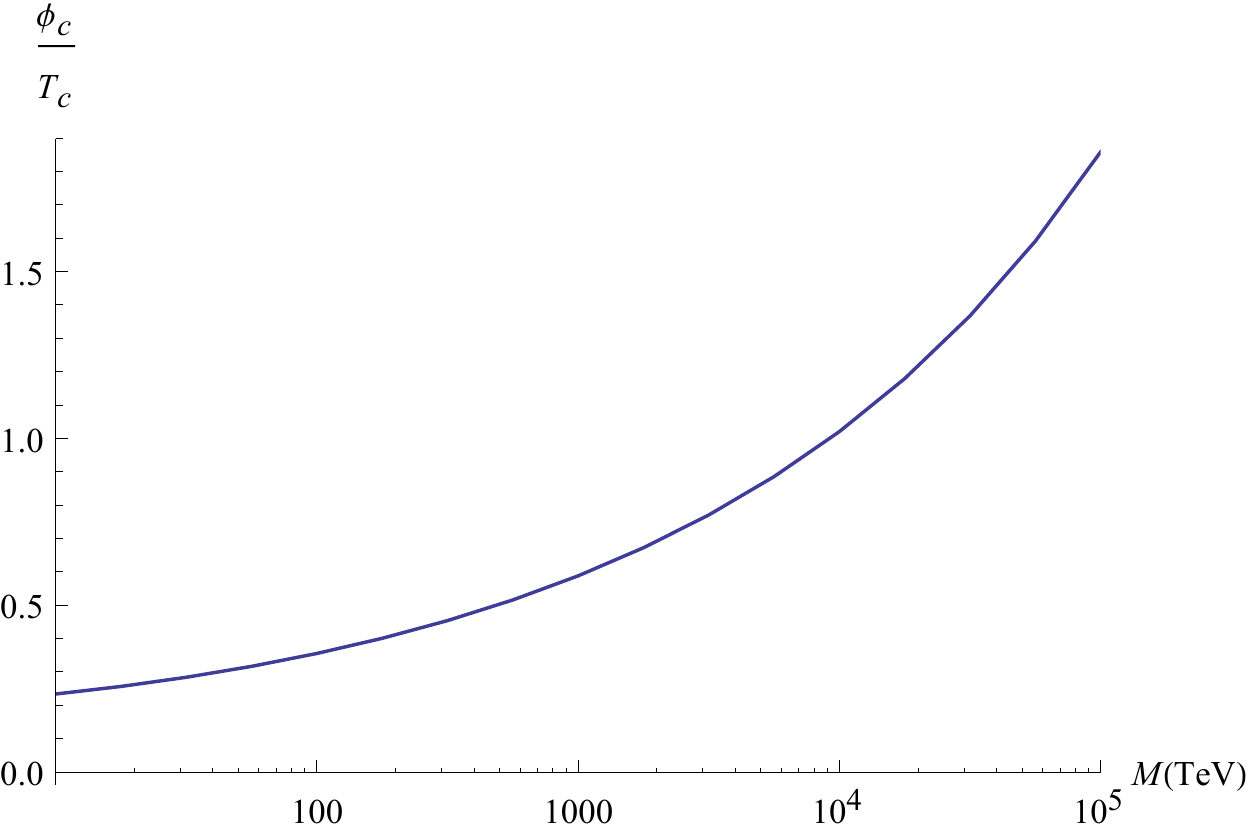}
\end{center}	 	
	 	\caption{The strength of the electroweak phase transition as a function of the monopole mass. Monopoles that are at least $9.3\cdot 10^3$ TeV in mass can satisfy the sphaleron washout condition and support baryogenesis.This corresponds to a Born-Infeld parameter of $\beta=1.6\cdot10^4(\text{TeV})^2$ \label{results}} 
	 \end{figure}

The production of monopoles cost in energy, so the free energy in the broken phase becomes:       
	 \begin{align}
	G_b=V(\phi_c(T))+n_0M
\label{bph}	 
	 \end{align}
Equating, now, the free energies in the symmetric and broken phases, it becomes clear that the electroweak phase transition happens at lower critical temperatures $T_c$, and, therefore, sphalerons may start to satisfy the non-washout condition (\ref{sphaleron}). As seen from Figure \ref{results}, this indeed takes place when monopoles are sufficiently heavy, $M> 9.3\cdot 10^3$ TeV. 

\subsection{Monopole - antimonopole annihilation  and monopole washout constraints}
The monopoles (antimonopoles) can drift towards antimonopoles (monopoles) through the scatterings on charged particles of plasma. Each of such scattering rate can be estimated as: $\sum_{q_i} \sigma_{q_iM}n_{q_i}=
\sum_{i}(hq_i/4\pi)^2 T^{-2}n_{q_i}=(3/4\pi^2)\zeta(3)T
\sum_{i}(hq_i/4\pi)^2$, where $\zeta(3)\approx 1.20$ and the sum goes over relativistic charged particles (we included only fermions) which are at thermal equilibrium at temperature $T$. After $\sim M/T$ such occurrences, a monopole will scatter at large angle and drift towards the antimonopole. Hence, the monopole/antimonopole mean free path is given by \cite{Preskill:1979zi}:
\begin{equation}
\lambda \approx \frac{M}{\sum_{q_i} \sigma_{q_iM}n_{q_i}T}\left(\frac{T}{M}\right)^{1/2}= 
\frac{1}{B}\left(\frac{M}{T^3}\right)^{1/2}
~,
\label{length}
\end{equation}
where $B=(3/4 \pi^2)\zeta(3)\sum_i(hq_i/ 4\pi)^2$. As long as the mean free path (\ref{length}) is smaller than the Coulomb capture distance $d_c$, monopole-antimonopole pairs can annihilate as described. However, as the universe expands and cools down, $\lambda$ grows faster than $d_c$, and below the temperature
\begin{equation}
T_f=\left(\frac{4\pi}{h^2}\right)^2\frac{M}{B^2}
\label{temp}
\end{equation}
where $\lambda\approx d_c$, the monopole-antimonopole annihilation rate becomes negligible. Solving the Boltzmann equation for monopole/antimonopole number density and evaluating it at $T_f$ one obtains\footnote{Here, we ignore the potential imbalance between the monopole and antimonopole number densities due to  CP violation.} \cite{Preskill:1979zi}:   
\begin{equation}
n_f=\frac{M}{Bh^2}\left(\frac{4\pi}{h^2}\right)^2\frac{T_f^3}{CM_P},
\label{nf}
\end{equation}
where $C=0.6N^{-1/2}$, $N$ is the number of relativistic degrees of freedom and $M_P$ is the Planck mass.  Below $T_f$  this number density simply dilutes as $T^{-3}$ due to the expansion of the universe. The monopole/antimonopole number density is constrained by the standard Big Bang nucleosynthesis. Namely, at $T=1$ MeV the monopole/antimonopole density should be such that:
\begin{align}
n/T^3=n_f/T_f^3\lesssim \frac{(1 \text{MeV})}{M}
\label{b1}
\end{align}
Plugging the numbers $B\approx 3$ and $C\approx 0.06$ and imposing an obvious requirement $T_f<T_c$, we obtain from (\ref{b1}) the following upper bound on the monopole mass:
\begin{align*}
M< 2.28\cdot 10^4~ \text{TeV}.
\end{align*} 
Hence, it is seen that the monopoles required to suppress sphalerons in the broken phase  can still satisfy the nucleosynthesis constraints. 

As has been mentioned above, quark/lepton scatterings off monopoles and antimonopoles also lead to unsuppressed anomalous $B+L$ violation \cite{Rubakov:1982fp}. These processes must also decouple once the equilibrium is achieved, otherwise they will washout the asymmetry. Hence, we demand the rate of such processes at $T_f$, $\sigma_{q_iM}n_f$, is less than the expansion rate of the universe, $H(T_f)=T_f^2/CM_P$,  which implies, 
\begin{align*}
M< \frac{4\pi BT_f}{\alpha^3}~,
\end{align*} 
 where $\alpha \approx 1/137$ is the fine structure constant. Taking into account Eq. (\ref{temp}), one immediately sees that the above inequality is always satisfied, without implying any extra constraint.

\section{Conclusion}

In this paper we have postulated the existence of electroweak monopoles regularized within the Born-Infeld hypercharge extension of the Standard Model. Such monopoles (and antimonopoles) must be copiously produced during the electroweak phase transition and can drive local nonequilibrium in plasma. The production of the monopoles cost in energy, thus postponing electroweak phase transition . We have shown that if monopole mass, defined through the Born-Infeld mass parameter, is within a narrow range $0.9 \cdot 10^4 ~ \text{TeV}<M<2.3\cdot 10^4 ~ \text{TeV}$, sphaleron mediated processes can be made ineffective, thus preventing washout of previously generated matter-antimatter asymmetry, while still satisfying the nucleosynthesis constraints. We have also verified that anomalous $B+L$ violation processes due to the quark/lepton - monopole scatterings, while being active during the phase transition at $\sim T_c$, become suppressed in the broken phase ($T<T_c$) due to the efficient enough monopole-antimonopole annihilation. 

If the electroweak phase transition is indeed accompanied by the production of the electroweak monopoles of mass, $M\sim 10^4 ~ \text{TeV}$, a new mechanism for the electroweak baryogenesis can be realised. Namely, out-of-equilibrium  quark/lepton scatterings off monopoles may generate non-zero $B+L$ number due to the anomaly. The issue of CP violation, which left outside this paper, must be considered carefully. We plan to study this mechanism in more detail in a future work.

\paragraph{Acknowledgement} We would like to thank Tsutomu Yanagida for useful discussions. This work was partially supported by the Australian Research Council.


\end{document}